\documentclass[11pt,eqs]{article}
\usepackage{latexsym}
\usepackage{bbold}
\usepackage{amsmath}
\usepackage{hyperref}
\hypersetup{
	colorlinks=true,
	linkcolor=cyan,
	filecolor=magenta,      
	urlcolor=cyan,
}

\makeatletter
\newcommand\xleftrightarrow[2][]{%
	\ext@arrow 9999{\longleftrightarrowfill@}{#1}{#2}}
\newcommand\longleftrightarrowfill@{%
	\arrowfill@\leftarrow\relbar\rightarrow}
\makeatother

\title{
	Simultaneous bosonic and fermionic T-dualization of the type II superstring theory - Buscher approach and double space representation
	\thanks{Work supported in part by the Serbian Ministry of Science, Technological Development and Innovation in the form of institutional financing. The data that support the findings of this study are openly available in arXiv at https://arxiv.org, reference number arXiv:2510.19536. There is no conflict of interest.}}
\author{B. Nikoli\'c \thanks{email: bnikolic@ipb.ac.rs}\\{\it Institute of Physics Belgrade, University of Belgrade, Pregrevica 118, Serbia} }

\begin{document}

\maketitle

\begin{abstract}
In this article I consider type II superstring in the pure spinor formulation with constant background fields in the context of T-dualization. First I prove that bosonic and fermionic T-dualization commute using already known T-dual transformation laws for bosonic and fermionic T-dualization. Consequently, the T-dual transformation laws of the full T-dualization are obtained. At the end the full T-dualization is realized in double space and it is showed that Buscher procedure and double space approach are equivalent in this specific case.
\end{abstract}

\section{Introduction}
\setcounter{equation}{0}

Duality is a feature emerged in string theory and relates two theories which are, at the first glance, different ones in the sense that they have different coupling constants and different spatial scales \cite{jopol}. It relates strong and weak couplings (S-duality), small and large dimensions (T-duality), making manifest their physical equivalence. T and S-dualities are transformations connecting five consistent superstring theories, so, generally, studying dualities is very important for better understanding of M-theory.

My focus in this article will be on T-duality transformations.
The simplest and most fundamental picture of T-duality transformation is a compactification of one spatial dimension on a circle of radius $R$. For closed string along compactified direction $X$ the following periodic condition must hold, $X(\sigma+2\pi)=X(\sigma)+2\pi R W$, where $W$ is the winding number. This periodic condition implies that momentum along that direction is quantizied, $P=K/R$, where $K$ is an integer named Kaluza-Klein excitation number. It turns out that theory, after compactification, becomes invariant under exchanges, $W\leftrightarrow K$ and $R\leftrightarrow \alpha'/R$, where $\alpha'$ is Regge slope parameter. In this way I show the physical equivalence of the theories on the large and small scales, which is the basic notion of T-duality.

Analytical form of the T-duality transformation is given by Buscher procedure \cite{buscher}, which is considered in literature as a definition of T-duality. The starting point of this approach is identifying the direction $X$ along which the global shift symmetry exists, $X\rightarrow X+a$. Further the observed symmetry is localized introducing covariant world-sheet derivative, $\partial_\pm X\rightarrow D_\pm X=\partial_\pm X+v_\pm$, where $v_\pm$ is a gauge field. Then the gauge condition is fixed putting $X=const.$. In order for the initial and transformed (T-dualized) theories to have the same number of degrees of freedom, one has to add a term with a Lagrange multiplier $y$ to the initial action. On the equation of motion for $y$ the initial theory is restored, while on the equations of motion for $v_\pm$ T-dual theory is obtained. One of the important outcomes of the procedure is analytical form of the relation between initial coordinates and T-dual coordinates - T-dual transformation laws. Buscher procedure is applicable on the models with constant backgrounds as well as on the cases where background fields do not depend on some coordinates \cite{nemci1,nemci2,nemci3,ameri,ilija,falk}.

The superstring theories have both bosonic and fermionic coordinates \cite{jopol}. Formally, Buscher procedure can be generalized for the fermionic coordinates and that is so called fermionic T-duality \cite{ferdual}. In the first paper of \cite{ferdual}, Berkovits and Maldacena showed that tree level superstring theories in presence of
supersymmetric background fields posses a new kind of symmetry. They analyzed the gluon scattering amplitudes in N = 4 super Yang-Mills theory and discovered that new symmetry - fermionic T-duality. It is a part of the dual
superconformal symmetry which should be connected to integrability and it is valid just at
string tree level.
Unlike bosonic T-duality, fermionic T-duality involves shifting along fermionic isometry directions. It is a non-local and formal operation, only defined at the level of the classical string sigma model and requiring a careful treatment of global and quantum consistency conditions.

Model used in this paper is type II superstring in pure spinor formulation \cite{pure}. In general, the model is pretty complex and all background fields and their field strengths depend on all coordinates $x^\mu$, $\theta^\alpha$ and $\bar\theta^\alpha$, where $\theta^\alpha$ and $\bar\theta^\alpha$ are fermionic coordinates. Because of that I make a strong approximation, in accordance with the consistency conditions, taking all background fields constant as well as Ramond-Ramond field strength.

The starting point are the papers \cite{radovi1} where bosonic and fermionic T-duality are analyzed separately in details using pure spinor formulation of the type II superstring theory with constant background fields. T-dual theory is obtained as well as T-dual transformation laws, which relates initial and T-dual coordinates. Also both bosonic and fermionic T-duality are represented in the double space \cite{doublesp}, space spanned by initial and T-dual coordinates (Lagrange multipliers). In double space T-dualization is realized through an appropriate permutation matrix and it acts as symmetry transformation. The equivalence of the Buscher analytical approach and algebraic double space approach for this specific choice of action is proved.

In this paper the intention is to analyze full T-dualized type II from \cite{radovi1} - theory after implementation of both fermionic and bosonic T-duality procedure. The first step in analysis is to check whether these two T-duality transformations commute. Commutation means that, independent of the sequence of T-dualizations, there are no new terms in the T-dual action (comparing to the initial one) and that all background fields are the same. In order to achieve that goal I use the results obtained in \cite{radovi1} - the form of the T-dual background fields and T-dual transformation laws. Combining these results it is showed that these two procedures commute for this specific choice of background fields. The action with constant background fields assumes a lot of approximations, so this check is not a trivial step in analysis. As a consequence of that conclusion, one obtained the T-dual transformation laws for full T-dualization as well as the form of the background fields of the full T-dualized theory.

Also the double space representation of the full T-dualization was done. The double space coordinate is of the form $Z^A=(x^\mu,\theta^\alpha,\bar\theta^\alpha, y_\mu, z_\alpha, \bar z_\alpha)^T$, where $x^\mu$, $\theta^\alpha$ and $\bar\theta^\alpha$ are the initial coordinates and $y_\mu$, $z_\alpha$ and  $\bar z_\alpha$ T-dual coordinates. Permutation matrix $\mathcal T^A{}_B$ is chosen in such a form to exchange the set of initial coordinates with the set of the T-dual coordinates and produces T-dual double coordinate, ${}^\star Z^A=\mathcal T^A{}_B Z^B$. Form of the action induces that there two generalized metrics, one for each chiral sector, $\mathcal H_+$ and $\mathcal H_-$. Demanding that T-dual double coordinate has the same T-dual transformation law as initial one, I get the form of the T-dual background fields and confirm the agreement with the analytical Buscher approach.

\section{Choice of action and brief review of the bosonic and fermionic T-duality}
\setcounter{equation}{0}

In this section the choice of the action will be justified. Further the brief review of the bosonic and fermionic T-duality of the chosen action will be given.

\subsection{Pure spinor formulation of type II superstring - justification of the specific choice of background fields}

The sigma model action for type II superstring in pure spinor formulation has the following general form
\begin{equation}\label{eq:dejstvo1}
S=S_0+V_{SG}\, ,
\end{equation}
where $S_0$ is the free theory action
\begin{equation}
S_0=\int_\Sigma d^2\xi \left( \frac{\kappa}{2}\eta^{mn}\eta_{\mu\nu}\partial_m x^\mu \partial_n x^\nu-\pi_\alpha \partial_{-} \theta^\alpha+\partial_+ \bar\theta^\alpha \bar\pi_\alpha\right)+S_\lambda+S_{\bar\lambda}\, ,
\end{equation}
while the second part is integrated vertex operator of the massless type II supergravity written in the condensed form as
\begin{equation}\label{eq:vsg}
V_{SG}=\int_\Sigma d^2 \xi (X^T)^M A_{MN}\bar X^N\, .
\end{equation}
Let me explain the notation and conventions. Here the vectors $X^M$ and $X^N$ are given as
\begin{equation}
X^M=\left(
\begin{array}{c}
\partial_+ \theta^\alpha\\\Pi_+^\mu\\d_\alpha\\\frac{1}{2}N_+^{\mu\nu}
\end{array}\right)\, ,\quad \bar X^M=\left(
\begin{array}{c}
\partial_-\bar\theta^\alpha\\\Pi_-^\mu\\\bar d_\alpha\\\frac{1}{2}\bar N_-^{\mu\nu}
\end{array}\right),
\end{equation}
and supermatrix $A_{MN}$, which components are background fields and field strengths, is of the form
\begin{equation}\label{eq:Amn}
A_{MN}=\left(\begin{array}{cccc}
A_{\alpha\beta} & A_{\alpha\nu} & E_\alpha{}^\beta & \Omega_{\alpha,\mu\nu}\\
A_{\mu\beta} & A_{\mu\nu} & \bar E_\mu^\beta & \Omega_{\mu,\nu\rho}\\
E^\alpha{}_\beta & E^\alpha_\nu & {\rm P}^{\alpha\beta} & C^\alpha{}_{\mu\nu}\\
\Omega_{\mu\nu,\beta} & \Omega_{\mu\nu,\rho} & \bar C_{\mu\nu}{}^\beta & S_{\mu\nu,\rho\sigma}
\end{array}\right)\, ,
\end{equation}
where notation and definitions are taken from Ref.\cite{verteks}. The fields $\lambda^\alpha$ and $\bar\lambda^\alpha$ are so called pure spinors, which actions, $S_\lambda$ and $S_{\bar\lambda}$, are free field ones, fully decoupled from the rest of action $S_0$. They will not be included in the further analysis.
The world sheet parameters are denoted by
$\xi^m=(\xi^0=\tau\, ,\xi^1=\sigma)$ and
$\partial_\pm=\partial_\tau\pm\partial_\sigma$. Bosonic  coordinates are $x^\mu$ ($\mu=0,1,2,\dots,9$) and $\theta^\alpha$ and $\bar\theta^{\alpha}$
$(\alpha=1,2,\dots,16)$ are fermionic coordinates. Their canonically conjugated momenta are $\pi_\alpha$ and $\bar
\pi_{\alpha}$. Number of space-time dimensions dictates the type of spinors - fermionic variables are Majorana-Weyl spinors and each of them has 16 independent real components. 

The first column and the first row of the matrix $A_{MN}$ contain auxiliary superfields, while
$A_{\mu\nu}$, $\bar E_\mu{}^\alpha$ and $E^\alpha{}_\mu$ are superfields and ${\rm P}^{\alpha\beta}$ is Ramond-Ramond field strength. All four make physical background. The fields, $\Omega_{\mu,\nu\rho}(\Omega_{\mu\nu,\rho})$, $C^\alpha{}_{\mu\nu}(\bar C_{\mu\nu}{}^\alpha)$ and $S_{\mu\nu,\rho\sigma}$, are field strengths for physical superfields. Much more details about derivation and structure of the action can be found in \cite{verteks}.

Implementation of the combined bosonic and fermionic T-duality is possible if there exists at least two isometry directions - one in the bosonic and one in fermionic sector. Consequently, one has to impose some restrictions. Trivial solution for bosonic T-dualization is to eliminate the explicit dependence on $x^\mu$ and keep only dependence on $\partial_\pm x^\mu$. Some specific models can contain $x^\mu$ dependence (see \cite{xzavisnost}) where isometry still exists. But here I chose the trivial solution which is in accordance withe the consistency conditions from Ref.\cite{verteks}.
For fermionic T-duality the situation is pretty similar. One has to avoid dependence on $\theta^\alpha$ and $\bar\theta^\alpha$. Such assumption produces the following consequence - the auxiliary superfields must be zero which is clear from the consistency conditions given in Ref.\cite{verteks}. Choosing physical background fields is not problematic in the sense of the consistency conditions \cite{verteks} except Ramond-Ramond (RR) field strength ${\rm P}^{\alpha\beta}$. There is a possibility of back reaction to the rest background fields. But detailed analysis presented in Ref.\cite{fermTd} (subsection 2.1) showed that choice of constant RR field strength is correct.

The quantities $d_\alpha$ and $\bar d_\alpha$ are generators of the BRST operators and they are supersymmetric combinations of the fermionic momenta and fermionic coordinates. Here is considered the model with constant background fields which means that $\theta^\alpha$ and $\bar\theta^\alpha$ dependence is neglected in the action. This choice of action implies that $d_\alpha\to \pi_\alpha$ and $\bar d_\alpha\to \bar\pi_\alpha$. Some additional arguments regarding the choice of the background fields and notation can be found in Ref.\cite{fermTd}.

After all, the final form of the type II superstring action in the pure spinor formulation is 
\begin{eqnarray}\label{eq:SB}
&{}&S=\kappa \int_\Sigma d^2\xi \left[\partial_{+}x^\mu
\Pi_{+\mu\nu}\partial_- x^\nu+\frac{1}{4\pi\kappa}\Phi R^{(2)}\right] \\&+&\int_\Sigma d^2 \xi \left[
-\pi_\alpha
\partial_-(\theta^\alpha+\Psi^\alpha_\mu
x^\mu)+\partial_+(\bar\theta^{\alpha}+\bar \Psi^{\alpha}_\mu
x^\mu)\bar\pi_{\alpha}+\frac{1}{2\kappa}\pi_\alpha P^{\alpha
\beta}\bar \pi_{\beta}\right ]\, .\nonumber
\end{eqnarray}
The terms with pure spinors disappeared because curvatures are zero for constant
physical superfields.

\subsection{Chiral gauge invariance}

Action for type II superstring theory given above (\ref{eq:SB}) is given in the form of the first order theory, because the fermionic canonical momenta are present in the action. For further analysis it is useful to use the equations of motion for $\pi_\alpha$ and $\bar\pi_\alpha$
\begin{equation}\label{eq:impulsi}
\pi_\alpha=-2\kappa \partial_+\left(\bar\theta^\beta+\bar\Psi^\beta_\mu x^\mu\right)(P^{-1})_{\beta\alpha}\, ,\quad \bar\pi_\alpha=2\kappa (P^{-1})_{\alpha\beta}\partial_-\left(\theta^\beta+\Psi^\beta_\mu x^\mu\right)\, ,
\end{equation}
and rewrite the action in the form of the second order theory
\begin{eqnarray}\label{eq:lcdejstvo}
&{}&S=\kappa \int_\Sigma d^2\xi \partial_+ x^\mu
\Pi_{+\mu\nu}\partial_-x^\nu\nonumber+\frac{1}{4\pi}\int_\Sigma d^2\xi \Phi R^{(2)}\\&+&2\kappa \int_\Sigma d^2\xi \partial_+\left(\bar\theta^\alpha+\bar\Psi^\alpha_\mu x^\mu\right)
(P^{-1})_{\alpha\beta}\partial_-\left(\theta^\beta+\Psi^\beta_\nu x^\nu\right)\nonumber\\
&=& \kappa \int_\Sigma d^2 \xi \partial_+ x^\mu \left[\Pi_{+\mu\nu}+2\bar\Psi^\alpha_\mu (P^{-1})_{\alpha\beta}\Psi^\beta_\nu\right]\partial_- x^\nu+\frac{1}{4\pi}\int_\Sigma d^2\xi \Phi R^{(2)}\\
&+&2\kappa \int_\Sigma d^2 \xi \left[\partial_+ \bar\theta^\alpha (P^{-1})_{\alpha\beta}\partial_-\theta^\beta +\partial_+ \bar\theta^\alpha (P^{-1}\Psi)_{\alpha\mu}\partial_-x^\nu+\partial_+x^\mu (\bar\Psi P^{-1})_{\mu\alpha}\partial_- \theta^\alpha\right]\nonumber\, .
\end{eqnarray}

In the obtained action it is obvious that $\bar\theta^\alpha$ goes only under $\partial_+$, while $\theta^\alpha$ goes under $\partial_-$. This means that one can make the transformation
\begin{equation}
\delta \theta^\alpha=\varepsilon^\alpha(\sigma^+)\, ,\quad \delta \bar\theta^\alpha=\bar\varepsilon^\alpha(\sigma^-)\, ,
\end{equation}
and the action stays unchanged. This is a kind of chiral symmetry. The complete analysis of the BRST approach in this case is given in Ref.\cite{fermTd}. Here I will make a review of the procedure as brief as it is possible. The standard notation and terminology are used.

The first step is introducing of the BRST transformations
\begin{equation}
s \theta^\alpha=c^\alpha(\sigma^+)\, ,\quad s \bar\theta^\alpha=\bar c^\alpha(\sigma^-)\, ,
\end{equation}
where $c^\alpha(\sigma^+)$ and $\bar c^\alpha(\sigma^-)$ are ghost fields, while $s$ is BRST nilpotent operator.
Gauge fixing procedure starts with so called gauge fermion
\begin{equation}
\Psi=2\kappa\int d^2\xi \left[\bar C_{\alpha}\left(\partial_+\theta^\alpha+\frac{\alpha^{\alpha\beta}}{2}b_{+\beta}\right)+\left(\partial_- \bar\theta^\alpha+\frac{1}{2}\bar b_{-\beta}\alpha^{\beta\alpha}\right)C_{\alpha}\right]\, ,
\end{equation}
where $\alpha^{\alpha\beta}$ is a non singular matrix,  $\bar C_{\alpha}$ and $C_{\alpha}$ are antighost fields, while $b_{+\alpha}$ and $\bar b_{-\alpha}$ are Nakanishi-Lautrup auxiliary fields. The latter satisfy the conditions
\begin{equation}
 s C_\alpha=b_{+\alpha}\, ,\quad s\bar C_\alpha=\bar b_{-\alpha}\, ,\quad s b_{+\alpha}=0\quad s \bar b_{-\alpha}=0\, .
\end{equation}
Acting with BRST transformation on gauge fermion produces the gauge fixed and Fadeev-Popov action
\begin{eqnarray}
&&s \Psi=S_{gf}+S_{FP}\nonumber \, ,\\
&& S_{gf}=2\kappa \int d^2\xi \left[\bar b_{-\alpha}\partial_+\theta^\alpha+\partial_-\bar\theta^\alpha b_{+\alpha}+\bar b_{-\alpha}\alpha^{\alpha\beta}b_{+\beta}\right]\nonumber\, ,\\
&&S_{FD}=2\kappa \int d^2\xi \left[\bar C_\alpha\partial_+c^\alpha+(\partial_-\bar c^\alpha) C_{\alpha}\right]\, .
\end{eqnarray}
It is obvious that Fadeev-Popov action is decoupled from the rest and it can be omitted in the further analysis.
On the equations of motion for $b$-fields
\begin{equation}
b_{+\alpha}=-(\alpha^{-1})_{\alpha\beta}\partial_+\theta^\alpha\, ,\quad \bar b_{-\alpha}=-\partial_-\bar\theta^\beta (\alpha^{-1})_{\beta\alpha}\, ,
\end{equation}
the BRST gauge fixed action gets its final form
\begin{equation}
S_{gf}=-2\kappa \int d^2\xi \partial_-\bar\theta^\alpha (\alpha^{-1})_{\alpha\beta}\partial_+\theta^\beta\, .
\end{equation}
The full action which is used in this analysis is the sum of the (\ref{eq:lcdejstvo}) and the BRST gauge fixed action
\begin{equation}\label{eq:sfull}
S_{full}=S+S_{gf}\, .
\end{equation}
In addition everything will be done through classical analysis, so, the T-dual transformation law for dilaton field will be omitted. Moreover, for constant dilaton field the term $\int \Phi R^{(2)}$ is a world-sheet Euler characteristic, which is a topological invariant. 

\subsection{A brief review of the bosonic T-duality}

Taking into account that in this subsection the main steps and results of the bosonic T-dualization procedure are represented, there is no need to work with $S_{full}$ (\ref{eq:sfull}) but with action $S$ (\ref{eq:lcdejstvo}) (omitting dilaton term).

The scheme of the Buscher T-dualization procedure assumes existence of, at least, one isometry direction. The action (\ref{eq:lcdejstvo}) is chosen in such a way that all background fields are constant and the theory is quadratic in the coordinate derivatives. This means that all bosonic (and fermionic) directions are isometry ones. So, all bosonic directions will be T-dualized simultaneously.
 
The world-sheet derivatives $\partial_\pm x^\mu$ are replaced with the appropriate covariant ones $D_\pm x^\mu=\partial_\pm x^\mu+v^\mu_\pm$, where $v_\pm^\mu$ are the gauge fields. Fixing the gauge, $x^\mu=const.$, the covariant derivative drives into gauge field. In order to make $v^\mu_\pm$ to be unphysical degrees of freedom, the term
\begin{equation}
S_{add}=\frac{\kappa}{2}\int d^2\xi (v_+^\mu \partial_- y_\mu-v_-^\mu \partial_+y_\mu)\, ,
\end{equation}
is added to the action, where $y_\mu$ are Lagrange multipliers (T-dual coordinates). Finally, the gauge fixed action is getting the form
{\small{\begin{eqnarray}\label{eq:sgf}
&{}&S_{fix}=S+S_{add}\\&=&\kappa\int d^2\xi \left[v_+^\mu \Pi_{+\mu\nu}v_-^\nu+2(\partial_+\bar\theta^\alpha+\bar\Psi^\alpha_\mu v^\mu_+)(P^{-1})_{\alpha\beta}(\partial_-\theta^\beta+\Psi^\beta_\nu v^\nu_-)+\frac{1}{2}(v_+^\mu \partial_- y_\mu-v_-^\mu \partial_+y_\mu)\right]\, .\nonumber
\end{eqnarray}}}

On the equations of motion for $y_\mu$
\begin{equation}\label{eq:jedy}
\partial_+ v_-^\mu-\partial_- v_+^\mu=0 \Rightarrow v_\pm^\mu=\partial_\pm x^\mu\, ,
\end{equation}
the gauge fixed action $S_{fix}$ transforms to the initial one, while on the equation of motion for gauge fields $v_\pm^\mu$ one gets
\begin{equation}
\check \Pi_{+\mu\nu}v^\nu_-+2\bar\Psi^\alpha{}_\mu (P^{-1})_{\alpha\beta}\partial_- \theta^\beta+\frac{1}{2}\partial_- y_\mu=0\, ,
\end{equation}
\begin{equation}
v^\nu_+ \check\Pi_{+\nu\mu}+2\partial_+\bar\theta^\alpha (P^{-1})_{\alpha\beta}\Psi^\beta{}_\mu-\frac{1}{2}\partial_+ y_\mu=0\, ,
\end{equation}
where
\begin{equation}
\check \Pi_{+\mu\nu}= \Pi_{+\mu\nu}+2\bar\Psi^\alpha{}_\mu (P^{-1})_{\alpha\beta}\Psi^\beta{}_\nu\equiv\check B_{\mu\nu}+\frac{1}{2}\check G_{\mu\nu}\, .
\end{equation}
From these equations it is possible to obtain expressions for gauge fields
\begin{equation}\label{eq:v-k}
v^\mu_-=-\kappa \check\Theta^{\mu\nu}_- \partial_-[y_\nu+4\bar\Psi^\alpha{}_\mu (P^{-1})_{\alpha\beta} \theta^\beta]\, ,
\end{equation}
\begin{equation}\label{eq:v+k}
v^\mu_+=\kappa \partial_+[y_\nu-4\bar\theta^\alpha(P^{-1})_{\alpha\beta}\Psi^\beta{}_\nu]\check\Theta^{\nu\mu}_-\, ,
\end{equation}
and inserting them into $S_{fix}$ one obtains the bosonic T-dual action
\begin{eqnarray}\label{eq:Sb}
&{}&{}^b S=\kappa \int d^2\xi \left[ \frac{\kappa}{2}\partial_+ y_\mu \check\Theta^{\mu\nu}_- \partial_-y_\nu+2\kappa \partial_+y_\mu \check\Theta^{\mu\nu}_- \bar\Psi^\alpha{}_\nu (P^{-1})_{\alpha\beta}\partial_-\theta^\beta\right.\label{eq:Sb} \\&-&\left.2\kappa \partial_+ \bar\theta^\alpha (P^{-1})_{\alpha\beta}\Psi^\beta{}_\mu \check\Theta^{\mu\nu}_- \partial_- y_\nu+2\partial_+\bar\theta^\alpha(P^{-1}-4\kappa P^{-1}\Psi \check\Theta_- \bar\Psi P^{-1})_{\alpha\beta}\partial_-\theta^\beta\right]\, .\nonumber
\end{eqnarray}
Here the top left index $b$ means bosonic T-dualized. Below is the clarification of the used notation
\begin{equation}\label{eq:tetapi1}
\check\Theta^{\mu\nu}_- \check\Pi_{+\nu\rho}=\frac{1}{2\kappa}\delta^\mu{}_\rho\, ,\quad \check\Theta^{\mu\nu}_-=-\frac{2}{\kappa}(\check G_E^{-1}\check\Pi_{-}\check G^{-1})^{\mu\nu}\, ,
\end{equation}
where
\begin{equation}
\check\Theta^{\mu\nu}_-=\Theta^{\mu\nu}_-- 4\kappa \Theta^{\mu\rho}_- \bar\Psi^\alpha{}_\rho (\tilde P^{-1})_{\alpha\beta} \Psi^\beta{}_\lambda \Theta^{\lambda\nu}_-=\check\Theta^{\mu\nu}+\frac{1}{\kappa}(\check G_E^{-1})^{\mu\nu}\, ,
\end{equation}
\begin{equation}\label{eq:tildeP}
\tilde P^{\alpha\beta}\equiv P^{\alpha\beta}+4\kappa\Psi^\alpha{}_\mu \Theta^{\mu\nu}_- \bar\Psi^\beta{}_\nu\, ,\quad \tilde P^{-1}_{\alpha\beta}=(P^{-1}-4\kappa P^{-1}\Psi \check\Theta_- \bar\Psi P^{-1})_{\alpha\beta}\, ,
\end{equation}
\begin{equation}
(\check G_E)_{\mu\nu}=\check G_{\mu\nu}-4 (\check B \check G^{-1} \check B)_{\mu\nu}\, ,
\end{equation}
\begin{equation}\label{eq:tetampip}
\Theta^{\mu\rho}_- \Pi_{+\rho\nu}=\frac{1}{2\kappa}\delta^\mu{}_\nu\, .
\end{equation}
The bosonic T-dualized action ${}^b S$ has the same form as the initial one (\ref{eq:lcdejstvo}), where $x^\mu$ is replaced by $y_\mu$, but with appropriate T-dual background fields. Comparing these two actions one gets their form
\begin{equation}\label{eq:bf1}
{}^b \Pi^{\mu\nu}_{+}=\frac{\kappa}{2}\Theta^{\mu\nu}_-\, ,
\end{equation}
\begin{equation}\label{eq:bf2}
{}^b \Psi^{\alpha\mu}=-\kappa \Psi^\alpha{}_\nu \Theta_-^{\nu\mu}\, ,\quad {}^b \bar\Psi^{\mu\alpha}=\kappa \Theta^{\mu\nu}_- \bar\Psi^\alpha{}_\nu\, ,
\end{equation}
\begin{equation}\label{eq:bf3}
{}^b P^{\alpha\beta}=\tilde P^{\alpha\beta}\, .
\end{equation}

Combining (\ref{eq:jedy}) with (\ref{eq:v-k}) and (\ref{eq:v+k}), the T-dual transformation laws are obtained
\begin{equation}\label{eq:tdlawx+}
\partial_+ x^\mu \cong \kappa  \partial_+\left( y_\nu-4\bar\theta^\alpha P^{-1}_{\alpha\beta}\Psi^\beta{}_\nu  \right)\check \Theta_-^{\nu\mu}\, ,
\end{equation}
\begin{equation}\label{eq:tdlawx-}
\partial_- x^\mu \cong -\kappa \check \Theta_-^{\mu\nu} \partial_-\left( y_\nu+4\bar\Psi^\alpha{}_\nu P^{-1}_{\alpha\beta} \theta^\beta\right)\, ,
\end{equation}
while inverse transformation laws are of the form
\begin{equation}\label{eq:tdlawy+}
\partial_+ y_\mu\cong -2\check \Pi_{-\mu\nu}\partial_+ x^\nu+4\partial_+ \bar\theta^\beta P^{-1}_{\beta\alpha}\Psi^\alpha_{\mu}  \, ,
\end{equation}
\begin{equation}\label{eq:tdlawy-}
\partial_- y_\mu\cong 2\partial_\pm x^\nu \check \Pi_{-\nu\mu}-4\bar\Psi^\alpha_{\mu} P^{-1}_{\alpha\beta} \partial_- \theta^\beta\, .
\end{equation}

\subsection{A brief review of the fermionic T-duality}

The mathematical procedure of the fermionic T-duality is the same as for the bosonic one. As it was said at the beginning of the previous subsection, all background fields are constant and the action $S_{full}$ (\ref{eq:sfull}) is quadratic in the coordinate derivatives, so, all fermionic directions are isometry ones. Let us repeat some important steps of the Buscher procedure along fermionic directions.

The first step is to introduce the covariant derivatives instead ordinary ones as well as the appropriate gauge fields
\begin{equation}
\partial_\pm\theta^\alpha \to D_\pm \theta^\alpha\equiv\partial_\pm\theta^\alpha+v_\pm^\alpha\, , \quad \partial_\pm\bar\theta^\alpha \to D_\pm
\bar\theta^\alpha\equiv\partial_\pm\bar\theta^\alpha+\bar v_\pm^\alpha\, .
\end{equation}
Further, the gauge is fixed, $\theta^\alpha=const.$ and $\bar\theta^\alpha=const.$, because of the existing shift symmetry along the fermionic directions. Initial and T-dual theory should have the same number of degrees of freedom, so, in order to make the gauge fields $v^\alpha_\pm$ and $\bar v^\alpha_\pm$ to be unphysical degrees of freedom, the terms with Lagrange multipliers have to be added to the full action $S_{full}$ (\ref{eq:sfull})
\begin{equation}
S_{add}=\frac{1}{2}\kappa \int_\Sigma d^2\xi \bar z_\alpha (\partial_+
v_-^\alpha-\partial_- v^\alpha_+)+\frac{1}{2}\kappa \int_\Sigma d^2\xi  (\partial_+
\bar v_-^\alpha-\partial_- \bar v^\alpha_+)z_\alpha\, ,
\end{equation}
where $z_\alpha$ and $\bar z_\alpha$ are grassmannian Lagrange multipliers.
Finally, the gauge fixed action is of the form
\begin{eqnarray}\label{eq:sgff}
&&S_{fix}=\kappa \int_\Sigma d^2\xi \partial_+ x^\mu \left[\Pi_{+\mu\nu}+2\bar\Psi^\alpha_\mu(P^{-1})_{\alpha\beta}\Psi^\beta_\nu\right]\partial_-x^\nu \\ &{}& +2\kappa \int_\Sigma \left[ \bar v_+^\alpha (P^{-1})_{\alpha\beta}v_-^\beta+\bar v_+^\alpha (P^{-1})_{\alpha\beta}\Psi^\beta_\nu\partial_-x^\nu+\partial_+x^\mu \bar\Psi^\alpha_\mu (P^{-1})_{\alpha\beta} v_-^\beta-\bar v_-^\alpha (\alpha^{-1})_{\alpha\beta}v_+^\beta\right]\nonumber \\ &{}& +\frac{\kappa}{2} \int_\Sigma d^2\xi \bar z_\alpha (\partial_+
v_-^\alpha-\partial_- v^\alpha_+)+\frac{\kappa}{2} \int_\Sigma d^2\xi  (\partial_+
\bar v_-^\alpha-\partial_- \bar v^\alpha_+)z_\alpha\, .\nonumber
\end{eqnarray}

The gauge fixed action (\ref{eq:sgff}) is a kind of auxiliary action - it produces both initial and T-dual action. Varying it with respect to the Lagrange multipliers, $z_\alpha$ and $\bar z_\alpha$, one gets
\begin{equation}\label{eq:jed3}
\partial_+
v^\alpha_--\partial_- v^\alpha_+=0 \Longrightarrow v_\pm^\alpha=\partial_\pm \theta^\alpha \, ,\quad \partial_+
\bar v^\alpha_--\partial_- \bar v^\alpha_+=0 \Longrightarrow \bar v_\pm^\alpha=\partial_\pm \bar\theta^\alpha\, ,
\end{equation}
which leads us to the initial action. On the other side, varying the gauge fixed action with respect to the gauge fields, $v^\alpha_\pm$ and $\bar v^\alpha_\pm$, the following set of equations is obtained
\begin{equation}\label{eq:jed1}
\bar v_-^\alpha=\frac{1}{4}\partial_- \bar z_\beta \alpha^{\beta\alpha}\, ,\quad \bar v_+^\alpha=\frac{1}{4}\partial_+ \bar z_\beta P^{\beta\alpha}-\partial_+ x^\mu \bar\Psi^\alpha_\mu\, ,
\end{equation}
\begin{equation}\label{eq:jed2}
v_+^\alpha=-\frac{1}{4}\alpha^{\alpha\beta}\partial_+ z_\beta\, ,\quad v_-^\alpha=-\frac{1}{4}P^{\alpha\beta}\partial_- z_\beta-\Psi^\alpha_\mu \partial_- x^\mu\, .
\end{equation}
Substituting these equations into the gauge fixed action (\ref{eq:sgff}), one becomes the expression for fermionic T-dual action
\begin{eqnarray}\label{eq:ftdual}
&{}& {}^f S=\kappa\int_\Sigma d^2\xi \partial_+ x^\mu  \Pi_{+\mu\nu}\partial_- x^\nu\, ,\\ &{}&+2\kappa\int_\Sigma d^2\xi\left[\partial_+\bar z_\alpha P^{\alpha\beta}\partial_- z_\beta -\partial_+x^\mu\bar\Psi^{\alpha}_{\mu} \partial_- z_\alpha+\partial_+\bar z_\alpha\Psi^{\alpha}_{\mu}\partial_- x^\mu-\partial_-\bar z_\alpha \alpha^{\alpha\beta} \partial_+ z_\beta\right]\, ,\nonumber
\end{eqnarray}
where the left subscript $f$ denotes fermionic T-dualization. The fermionic T-dual action has the same form as the initial one $S_{full}$ (\ref{eq:sfull}) up to the transition $\theta^\alpha\to z_\alpha$ and $\bar\theta^\alpha\to \bar z_\alpha$, with the corresponding T-dual background fields
\begin{equation}\label{eq:GBdual}
{}^f G_{\mu\nu}=G_{\mu\nu}+2\left[ (\bar\Psi P^{-1}\Psi)_{\mu\nu}+(\bar\Psi P^{-1}\Psi)_{\nu\mu}\right]\, ,\quad {}^f B_{\mu\nu}=B_{\mu\nu}+(\bar\Psi P^{-1}\Psi)_{\mu\nu}-(\bar\Psi P^{-1}\Psi)_{\nu\mu}\, ,
\end{equation}
\begin{equation}\label{eq:Psidual}
{}^f \Psi_{\alpha \mu}=4(P^{-1}\Psi)_{\alpha\mu}\, ,\; {}^f\bar\Psi_{\mu\alpha}=-4(\bar\Psi P^{-1})_{\mu\alpha}\, ,
\end{equation}
\begin{equation}\label{eq:Fdual}
{}^f P_{\alpha\beta}=16(P^{-1})_{\alpha\beta}\, ,\quad {}^f \alpha_{\alpha\beta}=16(\alpha^{-1})_{\alpha\beta}\, .
\end{equation}

Combining the Eq.(\ref{eq:jed3}) with the relations (\ref{eq:jed1}) and (\ref{eq:jed2}), one obtains the fermionic T-dual transformation laws
\begin{equation}\label{eq:ftlaw1}
\partial_-\theta^\alpha\cong-\frac{1}{4} P^{\alpha\beta}\partial_- z_\beta-\Psi^\alpha_\mu \partial_- x^\mu\, ,\quad
\partial_+\bar\theta^\alpha\cong\frac{1}{4}\partial_+ \bar z_\beta P^{\beta\alpha}-\partial_+ x^\mu \bar\Psi^\alpha_\mu\, ,
\end{equation}
\begin{equation}\label{eq:ftlaw2}
\partial_+\theta^\alpha\cong -\frac{1}{4}\alpha^{\alpha\beta}\partial_+ z_\beta\, ,\quad \partial_-\bar\theta^\alpha\cong \frac{1}{4}\partial_- \bar z_\beta \alpha^{\beta\alpha}\, ,
\end{equation}
while the inverse ones are
\begin{equation}\label{eq:inftlaw1}
\partial_- z_\alpha\cong -4(P^{-1})_{\alpha\beta}\partial_-\theta^\beta-4(P^{-1})_{\alpha\beta}\Psi^\beta_\mu\partial_- x^\mu\, ,\quad \partial_+\bar z_\alpha\cong 4\partial_+\bar\theta^\beta (P^{-1})_{\beta\alpha}+4\partial_+ x^\mu \bar\Psi^\beta_\mu (P^{-1})_{\beta\alpha}\, ,
\end{equation}
\begin{equation}\label{eq:invftlaw2}
\partial_+ z_\alpha\cong -4(\alpha^{-1})_{\alpha\beta}\partial_+\theta^\beta\, ,\quad \partial_- \bar z_\alpha\cong 4\partial_- \bar\theta^\beta (\alpha^{-1})_{\beta\alpha}\, .
\end{equation}

\section{Combined bosonic and fermionic T-dualization - full T-dualization}
\setcounter{equation}{0}

In this section the full T-dualization of type II superstring theory will be performed. It will be shown that bosonic and fermionic Buscher T-dualization procedures  commute for the action with the constant background fields (\ref{eq:sfull}). Proving that the full T-dual transformation laws will be obtained, and further, the background fields of the full T-dualized theory.

In order to prove commutation of these two procedures I will use the T-dual transformation laws, (\ref{eq:tdlawx+})-(\ref{eq:tdlawy-}) and (\ref{eq:ftlaw1})-(\ref{eq:invftlaw2}), as well as the expressions for the T-dual background fields, (\ref{eq:bf1})-(\ref{eq:bf3}) and (\ref{eq:GBdual})-(\ref{eq:Fdual}).

\subsection{First bosonic then fermionic T-dualization}

Let me consider the case where the bosonic T-dualization is done first and then the fermionic one. It is sufficient to consider only one subset of the T-dual transformation laws, for example, the relations where derivatives of the initial coordinates are expressed in terms of the derivatives of the undualized coordinates.

After bosonic T-dualization one has the the transformation laws
\begin{equation}\label{eq:dmx}
\partial_- x^\mu\cong -\kappa \check\Theta^{\mu\nu}_- \partial_- \left[ y_\nu+4\bar\Psi^\alpha{}_\nu (P^{-1})_{\alpha\beta}\theta^\beta\right]\, ,
\end{equation}
\begin{equation}\label{eq:dpx}
\partial_+ x^\mu\cong \kappa \partial_+ \left[ y_\nu-4\bar\theta^\alpha (P^{-1})_{\alpha\beta}\Psi^\beta{}_\nu\right] \check\Theta^{\nu\mu}_- \, ,
\end{equation}
and the theory has the background fields given in Eqs.(\ref{eq:bf1})-(\ref{eq:bf3}). Further the fermionic T-dualization procedure is performed starting from the action ${}^b S$ (\ref{eq:Sb}). The transformation laws (\ref{eq:ftlaw1}), taking into account (\ref{eq:bf1})-(\ref{eq:bf3}), are now of the form
\begin{equation}\label{eq:afterbdteta}
\partial_- \theta^\alpha \cong -\frac{1}{4}\tilde P^{\alpha\beta} \partial_- z_\beta-{}^b \Psi^{\alpha\mu} \partial_- y_\mu\, ,\quad \partial_+ \bar\theta^\alpha \cong \frac{1}{4}\partial_+ \bar z_\beta \tilde P^{\beta\alpha}-{}^b \bar\Psi^{\alpha\mu}\partial_+ y_\mu\, ,
\end{equation}
while transformation laws (\ref{eq:ftlaw2}) stay unchanged. Putting the first equation from (\ref{eq:afterbdteta}) into the Eq.(\ref{eq:dmx}) the following expression is obtained
\begin{equation}
\partial_- x^\mu\cong -\kappa \check \Theta^{\mu\nu}_- \partial_- y_\nu-4\kappa \check \Theta^{\mu\nu}_- \bar\Psi^\alpha{}_\nu (P^{-1})_{\alpha\beta} \left(-\frac{1}{4}\tilde P^{\beta\gamma}\partial_- z_\gamma-{}^b \Psi^{\beta\nu}\partial_-y_\nu\right)\, .
\end{equation}
Using the expression for bosonic T-dualized background field ${}^b \Psi^{\alpha\mu}$ (\ref{eq:bf2}) and the identity
\begin{equation}\label{eq:idty}
\check \Theta^{\mu\nu}_- \bar\Psi^\alpha{}_\nu (P^{-1})_{\alpha\beta}=\Theta^{\mu\nu}_- \bar\Psi^\alpha{}_\nu (\tilde P^{-1})_{\alpha\beta}\, ,
\end{equation}
the full T-dual transformation law is obtained in the form
\begin{equation}\label{eq:fullxm}
\partial_- x^\mu \cong -\kappa \Theta^{\mu\nu}_- \partial_- y_\nu+\kappa \Theta^{\mu\nu}_- \bar\Psi^\alpha_\nu \partial_- z_\alpha\, .
\end{equation}
Analogously, putting the second equation from Eq.(\ref{eq:afterbdteta}) into the Eq.(\ref{eq:dpx}) and using (\ref{eq:bf2}) and the mentioned identity (\ref{eq:idty}), the full T-dual transformation law for $\partial_+ x^\mu$ is obtained
\begin{equation}\label{eq:fullxp}
\partial_+ x^\mu \cong \kappa \partial_+ y_\nu \Theta^{\nu\mu}_- -\kappa \partial_+ \bar z_\beta \Psi^\beta{}_\nu \Theta^{\nu\mu}_-\, .
\end{equation}

The transformations given in Eqs.(\ref{eq:fullxm}), (\ref{eq:fullxp}), (\ref{eq:afterbdteta}) and (\ref{eq:ftlaw2}) are the T-dual transformation laws for full T-dualization of the type II superstring theory, where the bosonic T-dualization is performed first.

\subsection{First fermionic then bosonic T-dualization}

After performed fermionic T-dualization procedure there is a set of transformation laws given in Eqs.(\ref{eq:ftlaw1}), (\ref{eq:ftlaw2}), (\ref{eq:inftlaw1}) and (\ref{eq:invftlaw2}). The second in the row, the bosonic T-dualization, gives
\begin{equation}
\partial_+ x^\mu\cong \kappa \partial_+ \left[y_\nu-4\bar z_\alpha ({}^f P^{-1})^{\alpha\beta} \;{}^f \Psi_{\beta\nu}\right]{}^f \check\Theta^{\nu\mu}\, .
\end{equation}
Using the expressions for fermionic T-dualized background fields (\ref{eq:GBdual}), (\ref{eq:Psidual}) and (\ref{eq:Fdual}), the above equation gets the form
\begin{equation}\label{eq:fullxpfb}
\partial_+ x^\mu \cong \kappa\partial_+ y_\nu \Theta^{\nu\mu}_- - \kappa \partial_+ \bar z_\alpha \Psi^\alpha{}_\nu \Theta^{\nu\mu}_-\, .
\end{equation}
It is obvious the full T-dual transformation law for $\partial_+ x^\mu$ given in Eqs.(\ref{eq:fullxp}) and (\ref{eq:fullxpfb}) are the same, which suggests that bosonic and fermionic T-dualization in this specific case commute. For full confirmation it remains to check the rest of T-dual transformation laws.

For $\partial_- x^\mu$ the story is completely identical. After performed fermionic T-dualization, the transformation law (\ref{eq:dmx}) gets the form
\begin{equation}
\partial_- x^\mu\cong -\kappa {}^f \check\Theta^{\mu\nu}_- \partial_- \left[ y_\nu+4{}^f \bar\Psi_{\alpha\nu} ({}^f P^{-1})^{\alpha\beta} z_\beta \right]\, .
\end{equation}
Using (\ref{eq:GBdual}), (\ref{eq:Psidual}) and (\ref{eq:Fdual}), the above equation transforms into
\begin{equation}
\partial_- x^\mu\cong -\kappa \Theta^{\mu\nu}_- \partial_- y_\nu + \kappa \Theta^{\mu\nu}_- \bar\Psi^\alpha{}_\nu \partial_- z_\alpha\, .
\end{equation}
Inserting the obtained relations for $\partial_\pm x^\mu$ into expressions (\ref{eq:ftlaw1}), the expressions get the final form
\begin{equation}\label{eq:fullbtpfb}
\partial_+ \bar\theta^\alpha\cong \frac{1}{4}\partial_+ \bar z_\beta \tilde P^{\beta\alpha}-\kappa \partial_+ y_\mu\Theta^{\mu\nu}_- \bar\Psi^\alpha{}_\nu\, ,\quad \partial_- \theta^\alpha\cong -\frac{1}{4}\tilde P^{\alpha\beta}\partial_- z_\beta+\kappa \Psi^\alpha{}_\mu \Theta^{\mu\nu}_- \partial_- y_\nu\, .
\end{equation}
Taking into account the expressions for bosonic T-dualized gravitino fields (\ref{eq:Psidual}), the expressions same as in Eq.(\ref{eq:afterbdteta}) is obtained. The T-dual transformation laws after full T-dualization for $\partial_- \bar\theta^\alpha$ and $\partial_+ \theta^\alpha$ are given in Eqs.(\ref{eq:ftlaw2}) and (\ref{eq:invftlaw2}), because additional BRST gauge term is not affected by the bosonic T-dualization.

Finally, it is proved that bosonic and fermionic Buscher T-dualization procedures for this specific choice of background fields commute. Notice that commutation can be proved just on the level of the dualized background fields. But, in this way, I proved the commutation and, simultaneously, obtained the full T-dual transformation laws. Further it is easy to find the expressions for full T-dualized background fields
\begin{equation}\label{eq:bpjedn1}
{}^\star \Pi^{\mu\nu}_+=\frac{\kappa}{2}\check\Theta^{\mu\nu}_-\, ,
\end{equation}
\begin{equation}
{}^\star \Psi_\alpha{}^\mu=-4\kappa (P^{-1})_{\alpha\beta}\Psi^\beta{}_\nu \check\Theta^{\nu\mu}_-\, ,\quad {}^\star \bar\Psi^\mu{}_\alpha=-4\kappa \check\Theta^{\mu\nu}_- \bar\Psi^\beta{}_\nu (P^{-1})_{\beta\alpha}\, ,
\end{equation}
\begin{equation}\label{eq:bpjedn2}
{}^\star P_{\alpha\beta}=16 (\tilde P^{-1})_{\alpha\beta}\, ,\quad {}^\star \alpha_{\alpha\beta}=16 (\alpha^{-1})_{\alpha\beta}\, ,
\end{equation}
where $\star$ in the left superscript denotes full T-dualization.

\section{Double space representation of the full T-dualization}
\setcounter{equation}{0}

In the previous section I found the expressions for T-dual transformation laws in which the world-sheet derivatives of the initial coordinates $x^\mu$, $\theta^\alpha$ and $\bar\theta^\alpha$ are related with the derivatives of the coordinates of the full T-dualized theory $y_\mu$, $z_\alpha$ and $\bar z_\alpha$.
Here I want to rewrite the analytical T-dualization process in the double space. In order to accomplish that goal, first I have to find expressions for the inverse transformations.

\subsection{Inverse T-dual transformation laws for full T-dualization}

Let me start with the transformations for $\partial_+ x^\mu$ (\ref{eq:fullxpfb}) and $\partial_+ \bar\theta^\alpha$ (\ref{eq:fullbtpfb}) and multiply the first one with $\bar\Psi^\alpha{}_\mu$. Adding obtained expressions and using the definition of $\tilde P^{\alpha\beta}$ (\ref{eq:tildeP}), the following transformation is obtained
\begin{equation}
\partial_+ (\theta^\alpha+\bar\Psi^\alpha{}_\mu x^\mu)\cong \frac{1}{4}\partial_+ \bar z_\beta P^{\beta\alpha}\, ,
\end{equation}
and, consequently,
\begin{equation}
\partial_+ \bar z\alpha \cong 4\partial_+ (\theta^\beta+\bar\Psi^\beta{}_\mu)(P^{-1})_{\beta\alpha}\, .
\end{equation}
Analogously, it follows
\begin{equation}
\partial_- z_\alpha\cong -4(P^{-1})_{\alpha\beta}\partial_- (\theta^\beta+\Psi^\beta{}_\mu x^\mu)\, .
\end{equation}
Inserting these expressions into expressions for the full T-dual transformation law (\ref{eq:fullxpfb}) and multiplying by $\Pi_{+\mu\rho}$, the inverse relation is obtained
\begin{equation}\label{eq:fullyp}
\partial_+ y_\mu\cong 2\partial_+ x^\nu \check \Pi_{+\nu\mu}+4\partial_+ \bar\theta^\beta (P^{-1})_{\beta\alpha}\Psi^\alpha{}_\mu\, .
\end{equation}
Applying the same procedure on the opposite chiral sector, I get the other relation
\begin{equation}
\partial_- y_\mu \cong -2\check \Pi_{+\mu\nu}\partial_- x^\nu -4\bar\Psi^\alpha{}_\mu (P^{-1})_{\alpha\beta} \partial_- \theta^\beta\, .
\end{equation}
The rest two transformations fulfilling this set of transformations are given in Eqs.(\ref{eq:ftlaw2}) and (\ref{eq:invftlaw2}).

\subsection{Double space representation of the full T-dualization}

Buscher T-dualization procedure is an analytical procedure and it is considered in the literature as a definition of T-duality. But it is possible to make a matrix representation of the T-dualization introducing double space.

Double space was initially considered in the bosonic string case and it spanned by the double coordinates
\begin{equation}
Z^A=\left(
\begin{array}{c}
x^\mu\\
y_\mu
\end{array}
\right)\, ,
\end{equation}
where $y_\mu$ are the coordinates T-dual to the initial ones $x^\mu$. The T-dualization is represented by permutation matrix $\mathcal T^A{}_B$
\begin{equation}\label{eq:tauab}
\mathcal T^A{}_B=\left(
\begin{array}{cc}
0 & \mathbb{1}_D\\
\mathbb{1}_D & 0
\end{array}
\right)\, ,\quad
{}^\star Z^A=\mathcal T^A{}_B Z^B=\left(
\begin{array}{c}
y_\mu\\
x^\mu
\end{array}
\right)\, ,
\end{equation}
where $D$ denotes the dimension of the target space.
This is a full bosonic T-dualization but it is easy to generalize the approach to the case of the partial T-dualization (T-dualization of the subset of coordinates) \cite{parcijalna}. Further, it can be generalized to the fermionic T-dualization \cite{ferdual}. 

Let me rewrite the full T-dualization (bosonic+fermionic) in terms of the double space which is spanned by
\begin{equation}\label{eq:ZA}
Z^A=\left(
\begin{array}{c}
x^\mu\\
\theta^\alpha\\
\bar\theta^\alpha\\
y_\mu\\
z_\alpha\\
\bar z_\alpha
\end{array}\right)\, .
\end{equation}
From now on when I say double space coordinate I think on $Z^A$ given in (\ref{eq:ZA}). The full T-dualized double coordinate is defined as
\begin{equation}\label{eq:ZAfd}
{}^\star Z^A=\left(
\begin{array}{c}
y_\mu\\
z_\alpha\\
\bar z_\alpha\\
x^\mu\\
\theta^\alpha\\
\bar\theta^\alpha
\end{array}\right)\, ,
\end{equation}
where $\star$ denotes full T-dualization. The relation connection between these two double coordinates is realized by permutation matrix
\begin{equation}
{}^\star Z^A=\mathcal T^A{}_B Z^B\, ,
\end{equation}
where $\mathcal T^A{}_B$ is of the same form as in the bosonic case (\ref{eq:tauab}) but the identity operator is not $\mathbb{1}_{D\times D}$. It is now of the form
\begin{equation}\label{eq:id}
\mathbb{1}=\left(
\begin{array}{ccc}
\mathbb{1}_{D\times D} & 0 & 0\\
0 & \mathbb{1}_{16\times 16} & 0\\
0 & 0 & \mathbb{1}_{16\times 16}
\end{array}
\right)\, ,
\end{equation}
because the every fermionic variable is Majorana-Weyl spinor in $D=10$ and has 16 independent real components.

First I have to rewrite the T-dual transformation laws for full T-dualization in the appropriate form. Let me start with the T-dual law for $\partial_+ x^\mu$ (\ref{eq:fullxpfb}) and multiply from the right with
$$G^E_{\mu\nu}=-4\Pi_{\pm \mu \rho}(G^{-1})^{\rho\lambda}\Pi_{\mp \lambda\nu}\, .$$
As a result, the following equation is obtained
\begin{equation}
\partial_+ x^\mu G^E_{\mu\nu}\cong -2\partial_+ y_\rho (G^{-1}\Pi_-)^\rho{}_\nu+2\partial_+ \bar z_\beta \Psi^\beta{}_\rho (G^{-1}\Pi_-)^\rho{}_\nu\, ,
\end{equation}
where the definition of $\Theta^{\mu\nu}_-$ (\ref{eq:tetampip}) is used. Putting $\Pi_{-\mu\nu}=B_{\mu\nu}-\frac{1}{2}G_{\mu\nu}$ in the first term on the right-hand side of the equation, after short calculation, the following equation is obtained
\begin{equation}\label{eq:j1}
\partial_+ y_\mu \cong G^E_{\mu\nu}\partial_+ x^\nu -2(B G^{-1})_\mu{}^\nu \partial_+ y_\nu - 2(\Pi_+ G^{-1})_\mu{}^\nu \Psi^\alpha{}_\nu \partial_+ \bar z_\alpha\, .
\end{equation}
In the inverse relation (\ref{eq:fullyp}), using the same logic, one inserts
$$\check\Pi_{+\mu\nu}= \Pi_{+\mu\nu}+2\bar\Psi^\alpha{}_\mu (P^{-1})_{\alpha\beta}\Psi^\beta{}_\nu)\equiv\check B_{\mu\nu}+\frac{1}{2}\check G_{\mu\nu}\, ,$$
and at the end of calculation, I get
\begin{equation}\label{eq:j2}
\partial_+ x^\mu \cong 2(\check G^{-1}\check B)^\mu{}_\nu \partial_+ x^\nu+(\check G^{-1})^{\mu\nu}\partial_+ y_\nu+4(\check G^{-1})^{\mu\nu}\Psi^\alpha{}_\nu (P^{-1})^T_{\alpha\beta} \partial_+ \bar\theta^\beta\, .
\end{equation}
To complete the set of equations, I rewrite the already obtained equations in the slightly changed form
\begin{equation}\label{eq:j3}
\partial_+ \bar\theta^\alpha \cong \frac{1}{4}(\tilde P^T)^{\alpha\beta}\partial_+ \bar z_\beta+\kappa \bar\Psi^\alpha{}_\mu \Theta^{\mu\nu}_+ \partial_+ y_\nu\, ,
\end{equation}
\begin{equation}\label{eq:j4}
\partial_+ \bar z_\alpha \cong 4(P^{-1})^T_{\alpha\beta} \partial_+ \bar\theta^\beta+ 4(\bar\Psi P^{-1})_{\alpha\mu}\partial_+ x^\mu\, ,
\end{equation}
\begin{equation}\label{eq:j5}
\partial_+ \theta^\alpha\cong -\frac{1}{4}\alpha^{\alpha\beta}\partial_+ z_\beta\, ,\quad \partial_+ z_\alpha\cong -4(\alpha^{-1})_{\alpha\beta}\partial_+\theta^\beta\, .
\end{equation}
Note that $\Theta^{\mu\nu}_+$ is used. It is defined as $\Theta^{\mu\nu}_+= -\Theta^{\nu\mu}_-$.

In the same manner, the T-dual laws of the $-$ chiral sector are rewritten
\begin{equation}
-\partial_- y_\mu\cong G^E_{\mu\nu}\partial_-x^\nu-2(BG^{-1})_\mu{}^\nu \partial_- y_\nu + 2 (\Pi_- G^{-1}\bar\Psi)_\mu{}^\alpha \partial_- z_\alpha\, ,
\end{equation}
\begin{equation}
-\partial_- x^\mu \cong (\check G^{-1})^{\mu\nu}\partial_- y_\nu+2(\check G^{-1}\check B)^\mu{}_\nu \partial_- x^\nu + 4(\check G^{-1}\bar\Psi P^{-1})^\mu{}_\alpha \partial_- \theta^\alpha\, ,
\end{equation}
\begin{equation}
-\partial_- \theta^\alpha\cong \frac{1}{4}\tilde P^{\alpha\beta}\partial_- z_\beta-\kappa \Psi^\alpha{}_\mu \Theta^{\mu\nu}_- \partial_- y_\nu\, ,
\end{equation}
\begin{equation}
-\partial_- \bar\theta^\alpha\cong -\frac{1}{4}(\alpha^T)^{\alpha\beta} \partial_- \bar z_\beta\, ,
\end{equation}
\begin{equation}
-\partial_- z_\alpha\cong 4(P^{-1})_{\alpha\beta}\partial_- \theta^\beta+4(P^{-1}\Psi)_{\alpha\mu} \partial_- x^\mu\, ,
\end{equation}
\begin{equation}
-\partial_- \bar z_\alpha\cong -4 (\alpha^{-1})^T_{\alpha\beta}\partial_- \bar\theta^\beta\, .
\end{equation}

These two sets of $+$ and $-$ relations can be rewritten in the more condensed form
\begin{equation}
\pm \Omega_{MN} \partial_\pm Z^N\cong \mathcal{H}_{\pm MN} \partial_\pm Z^N\, ,
\end{equation}
where $\mathcal H_\pm$ is the generalized metric in the positive and negative chiral sector, while the matrix $\Omega$ 
\begin{equation}\label{eq:omega}
\Omega_{MN}=\left(
\begin{array}{cc}
0 & \mathbb{1}\\
\mathbb{1} & 0
\end{array}
\right)\, ,
\end{equation}
is known in double field theory (DFT) as invariant $SO(d,d)$ metric ($d=D+16+16$). Here $1$ is identity operator given in Eq.(\ref{eq:id}). Regarding the generalized metric $\mathcal H_\pm$, note that there are two metrics, for each chiral sector separately. They are of the form
\begin{equation}\label{eq:H+}
\mathcal H_{+MN}=\left(
\begin{array}{cccccc}
G^E & 0 & 0 & -2BG^{-1} & 0 & -2\Pi_+G^{-1}\Psi\\
0 & -4\alpha^{-1} & 0 & 0 & 0 & 0\\
4\bar\Psi P^{-1} & 0 & 4 (P^{-1})^T & 0 & 0 & 0\\
2\check G^{-1} \check B & 0 & 4 \check G^{-1}\Psi (P^{-1})^T & \check G^{-1} & 0 & 0\\
0 & 0 & 0 & 0 & -\frac{1}{4}\alpha & 0\\
0 & 0 & 0 & \kappa \bar\Psi \Theta_+ & 0 & \frac{1}{4}\tilde P^T
\end{array}
\right)\, ,
\end{equation}
\begin{equation}\label{eq:H-}
\mathcal H_{-MN}=\left(
\begin{array}{cccccc}
G^E & 0 & 0 & -2BG^{-1} & 2\Pi_- G^{-1}\bar\Psi & 0\\
4P^{-1} \Psi  & 4P^{-1} & 0 & 0 & 0 & 0\\
0 & 0 & -4 (\alpha^{-1})^T & 0 & 0 & 0\\
2\check G^{-1} \check B & 4 \check G^{-1}\bar\Psi P^{-1} & 0 & \check G^{-1} & 0 & 0\\
0 & 0 & 0 & -\kappa \Psi \Theta_- & \frac{1}{4}\tilde P & 0\\
0 & 0 & 0 & 0 & 0 & -\frac{1}{4}\alpha^T
\end{array}
\right)\, ,
\end{equation}
where the space-time ($\mu,\nu,\rho,\dots$) and fermionic indices ($\alpha,\beta,\gamma,\dots$) are omitted for practical reasons. As it is in the pure bosonic case, the matrices $\mathcal H_\pm$ and $\Omega$ satisfy the following conditions
\begin{equation}
\mathcal H_\pm^T \Omega \mathcal H_\pm=\Omega\, ,\quad \Omega^2=1\, ,\quad s\det \mathcal H_\pm=1\, .
\end{equation}
Here $s\det$ denotes superdeterminant, because $\mathcal H_\pm$ contains four blocks which are supermatrices.

The basic demand is that T-dual transformation law for the T-dual coordinate ${}^\star Z^A$ has the same form as the T-dual transformation law for $Z^A$
\begin{equation}
\pm \Omega_{MN} \partial_\pm {}^\star Z^N\cong {}^\star \mathcal{H}_{\pm MN} \partial_\pm {}^\star Z^N\, .
\end{equation}
Consequently, T-dual generalised metric, $\mathcal H_{\pm MN}$, transform as
\begin{equation}\label{eq:Hpmdual}
{}^\star \mathcal H_{\pm MN}=\mathcal T_M{}^P \mathcal H_{\pm PQ} \mathcal T^Q{}_N\, . 
\end{equation}
From this equation, it is possible to find the background fields of the full T-dualized theory. The goal is to compare them with those obtained using analytical method and confirm applicability of double space formalism in this case.

Let the $+$ chiral sector be the first for analysis. Taking into account that matrix (\ref{eq:H+}) is big one, it is simplified as
\begin{equation}
\mathcal H_{+MN}=\left(
\begin{array}{cc}
A & B\\
C & D
\end{array}
\right)\, ,
\end{equation}
where these $A$, $B$, $C$ and $D$ are $(D+32)\times(D+32)$ matrices. Using the relation (\ref{eq:Hpmdual}) and this block structure of the matrix $\mathcal H_{+MN}$, the following result is obtained
\begin{equation}
{}^\star \mathcal H_{+MN}=\left(
\begin{array}{cc}
D & C\\
B & A
\end{array}
\right)\, .
\end{equation}
Because T-dualized theory is physically equivalent to the initial one, the structure of the ${}^\star\mathcal H_{+MN}$ is of the form
\begin{equation}
{}^\star \mathcal H_{+MN}=\left(
\begin{array}{cc}
{}^\star A & {}^\star B\\
{}^\star C & {}^\star D
\end{array}
\right)\, .
\end{equation}
Comparing the right-hand sides of the last two equations, from ${}^\star A=D$, the following expressions for T-dual background fields are obtained
\begin{equation}\label{eq:jedn1}
{}^\star (G^E)^{\mu\nu}=(\check G^{-1})^{\mu\nu}\, ,
\end{equation}
\begin{equation}
-4{}^\star (\alpha^{-1})^{\alpha\beta}=-\frac{1}{4}\alpha^{\alpha\beta} \Longrightarrow {}^\star \alpha_{\alpha\beta}=16 (\alpha^{-1})_{\alpha\beta}\, .
\end{equation}
\begin{equation}
4[{}^\star (P^{-1})^T]^{\alpha\beta}=\frac{1}{4}(\tilde P^T)^{\alpha\beta}\Longrightarrow {}^\star P_{\alpha\beta}=16\tilde P^{-1}_{\alpha\beta}
\end{equation}
\begin{equation}
4{}^\star \bar\Psi^\mu{}_\beta ({}^\star P^{-1})^{\beta\alpha}=-\kappa \Theta_-^{\mu\nu}\bar\Psi_\nu{}^\alpha \Longrightarrow {}^\star \bar\Psi^\mu{}_\alpha=-4\kappa \Theta^{\mu\nu} _- \bar\Psi_\nu{}^\beta (\tilde P^{-1})_{\beta\alpha}\, .
\end{equation}
From the equality of block components ${}^\star B=C$, it follows
\begin{equation}\label{eq:jedn2}
-2 {}^\star B^{\mu\rho}({}^\star G^{-1})_{\rho\nu}=2 (\check G^{-1})\check B_{\rho\nu}\, .
\end{equation}
Using the simple fact that $\frac{1}{2}{}^\star G {}^\star G^{-1}=\frac{1}{2}\check G^{-1} \check G$, I get the following useful relation
\begin{equation}\label{eq:auxrel1}
{}^\star \Pi_{\pm}^{\mu\rho} ({}^\star G^{-1})_{\rho\nu}=-(\check G^{-1})^{\mu\rho}\check \Pi_{\mp\rho\nu}\, .
\end{equation}
Further, from ${}^\star B=C$, it follows
\begin{equation}
-2 {}^\star \Pi_+^{\mu\rho}({}^\star G^{-1})_{\rho\lambda} {}^\star \Psi^\lambda{}_\alpha=4(\check G^{-1})^{\mu\rho}\Psi_\rho{}^\beta (P^{-1})^T_{\beta\alpha}\, ,
\end{equation}
and using (\ref{eq:auxrel1}) and assumed non singularity of $\check G_{\mu\nu}$, the final expression for ${}^\star \Psi$ is obtained
\begin{equation}
{}^\star \Psi^\mu{}_\alpha=4\kappa \check\Theta_+^{\mu\nu}P^{-1}_{\alpha\beta} \Psi^\beta{}_\mu = -4\kappa P^{-1}_{\alpha\beta} \Psi^\beta{}_\nu \check\Theta^{\nu\mu}_-\, .
\end{equation}
Here the fact $\check\Theta_+=-\check \Theta_-^T$ is used.

Let me go back to the equation (\ref{eq:jedn1}). The T-dual effective metric ${}^\star G_E^{\mu\nu}$ is defined as
\begin{equation}
{}^\star G_E^{\mu\nu}= -4{}^\star \Pi_+^{\mu\rho}({}^\star G^{-1})_{\rho\lambda}{}^\star \Pi_-^{\lambda\nu}\, .
\end{equation}
Now using the auxiliary relation (\ref{eq:auxrel1}), one gets
\begin{equation}
{}^\star G_E^{\mu\nu}=-4(\check G^{-1})^{\mu\rho}\check \Pi_{-\rho\lambda} (\check G^{-1})^{\lambda\sigma}\check \Pi_{+\sigma\omega}{}^\star G^{\omega\nu}=(\check G^{-1})^{\mu\rho}\check G^E_{\rho\lambda}{}^\star G^{\lambda\nu}\, .
\end{equation}
At the end, using (\ref{eq:jedn1}), finally one obtains the result
\begin{equation}
{}^\star G^{\mu\nu}=(\check G_E^{-1})^{\mu\nu}\, .
\end{equation}
As the undetermined T-dual background field remained ${}^\star B^{\mu\nu}$. Using expression for ${}^\star G^{\mu\nu}$ and relation (\ref{eq:jedn2}), I get
\begin{equation}
{}^\star B^{\mu\rho} \check G^E_{\rho\nu}= - (\check G^{-1})^{\mu\rho}\check B_{\rho\nu}\Longrightarrow {}^\star B^{\mu\nu}=-(\check G^{-1}\check B \check G_E^{-1})^{\mu\nu}=\frac{\kappa}{2}\check\Theta^{\mu\nu}\, .
\end{equation}
Finally
\begin{equation}
{}^\star \Pi_{+}^{\mu\nu}={}^\star B^{\mu\nu}+\frac{1}{2}{}^\star G^{\mu\nu}=\frac{\kappa}{2}\check \Theta^{\mu\nu}_-\, .
\end{equation}

Comparing T-dual background fields obtained in the analytical approach (\ref{eq:bpjedn1})-(\ref{eq:bpjedn2}) with the expressions above obtained in the double space approach, the full accordance is obvious. The same expressions would be obtained analyzing $\mathcal H_{-MN}$.

\section{Concluding remarks}
\setcounter{equation}{0}

In this paper the pure spinor model of type II superstring theory is studied. I choose to work with the superstring propagating in the presence of the constant background fields. The context of studying is T-dualization - combined bosonic and fermionic T-duality and double space representation of the full (bosonic$+$fermionic) T-dualization. 

At the beginning there are the separate recapitulations of the known results for bosonic and fermionic T-dualization of the type II superstring theory with constant background fields. The short discussion about choice of background fields is given and it is shown briefly the accordance with the consistency conditions. The main steps and results from \cite{radovi1} are repeated. Also in these papers \cite{radovi1} one can find more detailed analysis regarding the choice of the background fields.

In the next part of the paper I showed that bosonic and fermionic Buscher T-dualization procedure, applied on the type II superstring with constant background fields, commute. Independent on sequence of the application of T-duality procedures, the result is unique - full T-dualized theory. Further, the background fields of the full T-dualized theory are obtained. The result is not trivial in the sense that the model analyzed in his paper is approximated - all background fields are constant. This result is an additional proof that such choice is correct.

Final part of the paper is dedicated to the representation of full T-dualization in double space. Here I introduced the double coordinate and define the matrix of the full T-dualization, $\mathcal T^M{}_N$, which exchanges $x^\mu\leftrightarrow y_\mu$, $\theta^\alpha\leftrightarrow z_\alpha$ and $\bar\theta^\alpha\leftrightarrow \bar z_\alpha$, simultaneously. The full T-dualized coordinate is defined as ${}^\star Z^M=\mathcal T^M{}_N Z^N$. The next step was to rewrite the T-dual transformation laws using double coordinate and introducing so called generalized metric $\mathcal H_{\pm MN}$. Note that every chiral sector has its own generalized metric. Demanding that initial double coordinate $Z^M$ and fully T-dualized ${}^\star Z^M$ have the same T-dual transformation laws I get the form of the T-dualized generalized metric, ${}^\star\mathcal H_\pm=\mathcal T^T \mathcal H_\pm \mathcal T$. From one of these two equations I calculate the background fields of the fully T-dualized theory which are in full accordance with those obtained in the analytical (Buscher) procedure. For this specific choice of background fields I proved the equivalence of the analytical and double space approach.

At the end it is interesting to make comparison of the results from this paper with the recently published ones regarding double superspaces, double field theory and T-dualities. In the paper \cite{dodatak1} it is proved that abelian Yang-Baxter (YB) deformations are equivalent to the commuting TsT transformations (T-dualities and coordinate shifts). Further, the result is generalised to fermionic deformations and fermionic T-duality, and finally to the super T-duality group $OSp(d_b,d_b|2d_f)$. This is in full correspondence with the result presented in this paper. On the other hand, Refs.\cite{dodatak2,dodatak3} considers the non-Abelian T-duality and its supersymmetric generalization. It is concluded taht there is a problem of integrating out all the initial variables in general case but in the case of the principal $OSp(1,2)$ model this problem disappears - super T-dualities commutes as it is the conclusion of this paper. In Refs.\cite{dodatak4,dodatak5,dodatak6} it is considered the double field theory with the $OSp(d_b,d_b|2d_f)$ symmetry. As it is proved in \cite{dodatak1}, super YB deformations are equivalent to the commuting super T-dualizations.

\end{document}